\documentclass[11pt,preprint]{elsarticle}
\usepackage[utf8]{inputenc}
\usepackage{cancel}
\usepackage{graphicx}
\usepackage{epsfig}
\usepackage{url}
\usepackage{epstopdf}
\usepackage[caption=false]{subfig}
\pdfoutput=1
\usepackage[english]{babel}
\usepackage[T1]{fontenc}
\usepackage{amsmath}
\usepackage{booktabs}
\usepackage{pifont}
\usepackage{hyperref}
\usepackage{bm}
\usepackage{amsmath, amssymb}
\usepackage{tikz}
\usetikzlibrary{positioning, shapes, arrows}
\usepackage{setspace} 
\newtheorem{theorem}{Theorem}

\begin{document}
\makeatletter
\def\ps@pprintTitle{%
  \let\@oddhead\@empty
  \let\@evenhead\@empty
  \def\@oddfoot{}% \hfill\today se quiser manter a data, ou vazio para limpar tudo
  \let\@evenfoot\@oddfoot}
\makeatother
\begin{frontmatter}

\title{Modular Aggregation as a Debiasing Method for Non-Stationary Discrete Sources: Convergence and Numerical Validation}

\author{Eduardo Gueron\corref{cor1}}
\ead{eduardo.gueron@ufabc.edu.br}

\address{Centro de Matem\'{a}tica Computa\c{c}\~{a}o e Cogni\c{c}\~{a}o -- Universidade Federal do ABC, Santo Andr\'{e}, Brazil}


\begin{abstract}
We analyze \emph{modular aggregation}---summing $N$ independent outcomes
modulo $m$---as a post-processing method for extracting nearly uniform
randomness from biased discrete sources. Using discrete Fourier analysis over
the cyclic group $\mathbb{Z}_m$, we prove exponential convergence of the
output distribution to uniformity, with a rate determined by the largest
non-trivial Fourier modulus. The result applies to independent non-stationary
(non-IID) sources under a uniform spectral-gap condition on the non-trivial
Fourier modes. Numerical simulations under several bias regimes, including
cyclic drift and extreme cyclic bias, are used as finite-sample diagnostics
and illustrate the theoretical predictions in comparison with Peres extraction
and SHA-256 post-processing. The robustness of modular aggregation comes at a
retention cost of order $1/N$, yielding an explicit trade-off between
statistical quality and throughput.
\end{abstract}


\end{frontmatter}

% -
\section{Introduction}
\label{sec:introduction}
% -

Extracting uniform randomness from biased physical sources is a recurring
problem in Monte Carlo simulation, random-number generation, and stochastic
modelling~\cite{Preskill2018, Woo2024, Haw2015}. Because physical devices are
never perfectly unbiased, post-processing is often required to transform raw,
skewed outputs into statistically uniform sequences. The classical procedure
due to Von Neumann~\cite{vonNeumann1951}, later extended by
Peres~\cite{Peres1992} and by tree-based algorithms~\cite{StoutWarren1984},
achieves this by discarding suitable patterns in the input sequence. A
parallel line of work in theoretical computer science has developed the
general theory of randomness extractors~\cite{Vadhan2012, Chattopadhyay2019},
while Markov-chain methods have also been proposed~\cite{Morina2022}. In
practice, cryptographic hash functions such as SHA-256~\cite{NIST2015} are
also widely used as post-processing tools.

A common assumption in much of this literature is that the source is either
stationary, with identically distributed draws, or subject to stochastic
fluctuations without a persistent deterministic structure. Many real devices,
however, may exhibit \emph{structured} non-stationarity: slow drifts arising
from thermal gradients, ageing effects, feedback mechanisms, or periodic duty
cycles. The behaviour of standard post-processing methods under such
conditions is more delicate, especially when the bias evolves coherently over
time. This issue is the main motivation for the present work.

We study \emph{modular aggregation}: given $N$ independent outcomes $r_1,\ldots,r_N$ from a discrete source over $\{0,1,\ldots,m-1\}$, the
processed output is $r_f=\sum_{j=1}^{N} r_j \pmod{m}$.
% deixei inline a equação
This operation can be interpreted as a random walk on the cyclic group
$\mathbb{Z}_m$, a perspective developed in the probabilistic literature on
random walks on groups~\cite{Diaconis1988, aldousfill2002,
DiaconisFreedman1999}. In this setting, convergence to uniformity is governed
by the non-trivial Fourier modes of the increment distributions.

Our contribution is twofold. First, we give an explicit Fourier-based proof
of exponential convergence to uniformity for independent, possibly
non-stationary sources, under a uniform spectral-gap condition on the
non-trivial Fourier coefficients. The resulting total-variation bound makes
the dependence on the aggregation size $N$ and on the largest non-trivial
Fourier modulus explicit (Section~\ref{sec:theory}). Second, we present
numerical simulations under several bias regimes, including cyclic drift and
extreme cyclic bias, comparing modular aggregation with Peres extraction and
SHA-256 post-processing (Section~\ref{sec:numerical_validation}). These
statistical tests are used as finite-sample diagnostics; the analytical
guarantee concerns the distribution of the modular output on
$\mathbb{Z}_m$.

The robustness of modular aggregation comes at a transparent cost: only one
output symbol is produced from each block of $N$ raw symbols, so the retention
rate scales as $1/N$. The method therefore provides an explicit trade-off
between statistical quality and throughput. This trade-off is particularly
visible in highly biased regimes, where the spectral gap is small and a larger
aggregation size is required.

% -
\section{Method: Modular Aggregation}
\label{sec:method}
% -

Consider a source producing outcomes in $\{0, 1, \ldots, m-1\}$. Let
$p_k^{(j)}$ denote the probability of outcome $k$ at trial $j$. Given $N$
independent samples $r_1, r_2, \ldots, r_N$, the modularly aggregated output
is
\begin{equation}
  r_f \;\equiv\; \sum_{j=1}^{N} r_j \pmod{m}.
  \label{eq:modular_sum}
\end{equation}
The goal is to show that the distribution of $r_f$ approaches uniformity as
$N$ grows.  What matters is the spectral quantity
$|G_i(\omega^{j})|$: convergence is guaranteed whenever all non-trivial
Fourier moduli remain uniformly below one (see Theorem~\ref{theo:main}).
Full support is a simple sufficient condition for each individual
non-trivial Fourier modulus to be strictly smaller than one, but it is not
the relevant uniform condition in the non-stationary setting. Moreover,
support generation alone is not sufficient; for instance, the deterministic
choice $r_i\equiv 1$ has support that generates $\mathbb{Z}_m$ when
$\gcd(1,m)=1$, yet the modular sum remains deterministic and never converges
to uniformity.

% -
\section{Theoretical Convergence}
\label{sec:theory}
% -

We establish convergence using probability generating functions (PGFs) and the
discrete Fourier transform on $\mathbb{Z}_m$~\cite{Diaconis1988}. Let $r_i$ be an
independent random variable representing the outcome at trial $i$, with
distribution $P(r_i = k) = p_k^{(i)}$ on $\{0, \ldots, m-1\}$ and PGF $G_i(t)$.
For the partial sum $S_N = r_1 + \cdots + r_N$, the PGF factorises as
$G_{S_N}(t) = \prod_{i=1}^{N} G_i(t)$.

Setting $\omega = e^{2\pi i/m}$ and using the orthogonality of characters
of $\mathbb{Z}_m$, the modular probability can be written as:
\begin{equation}
  P(r_f \equiv k \pmod{m})
  = \frac{1}{m}\sum_{j=0}^{m-1} \omega^{-jk} \prod_{i=1}^{N} G_i(\omega^j).
  \label{eq:prob_k_nonstat}
\end{equation}

The formal bound on the deviation from uniformity is established by the
following result regarding the total variation distance.

\begin{theorem}\label{theo:main}
Let $(r_i)_{i\ge 1}$ be independent random variables on $\mathbb{Z}_m$, and set
\[
\rho_i \;:=\; \max_{j\neq 0}|G_i(\omega^{j})|, 
\qquad 
\rho \;:=\; \sup_{i\ge 1}\rho_i < 1.
\]
(Equivalently, every non-trivial Fourier coefficient is uniformly bounded in modulus by the constant $\rho$.)  
Then the total-variation distance between the distribution of the modular sum $S_N\bmod m$ and the uniform distribution $\mathcal{U}$ on $\mathbb{Z}_m$ satisfies
\begin{equation}
\bigl\| P_{S_N}-\mathcal{U}\bigr\|_{\mathrm{TV}}
\;\le\; \frac{m-1}{2}\;\prod_{i=1}^{N}\rho_i
\;\le\; \frac{m-1}{2}\,\rho^{N}.
\label{eq:error_bound}
\end{equation}
\end{theorem}

\textit{Proof.} From Eq.~\eqref{eq:prob_k_nonstat}, the deviation from uniformity at any $k$ is bounded by $\frac{1}{m}\sum_{j=1}^{m-1} \prod_{i=1}^N |G_i(\omega^j)|$. Using the definition of total variation distance, $\| P_{S_N} - \mathcal{U} \|_{\mathrm{TV}} = \frac{1}{2} \sum_{k=0}^{m-1} |P(r_f \equiv k \pmod{m}) - \frac{1}{m}|$, the uniform bound $|G_i(\omega^j)| \le \rho_i$ for $j \ne 0$ yields Eq.~\eqref{eq:error_bound}. Since $\rho < 1$ by assumption, the product vanishes exponentially as $N \to \infty$.\hfill$\square$

\paragraph{Stationary Case (IID) and Numerical Example}
For a stationary source, $G_i(t) = G(t)$ and $\rho_i = \rho$ for all $i$, making the bound exactly $\tfrac{m-1}{2}\rho^N$. For example, with $m = 6$, $p_0 = 0.8$, and $p_k = 0.04$ for $k > 0$, direct computation gives $\rho \approx 0.76$. The statistical distance decays as $0.76^N$, falling below $10^{-10}$ for $N \approx 80$. In adversarial extreme-bias scenarios (e.g., dominant probability 0.99), a much larger $N$ is required, as quantified in Section~\ref{sec:numerical_validation}.
%--  
\section{Numerical Validation}
\label{sec:numerical_validation}

We conducted two sets of simulations: (i) entropy convergence across
representative source models, and (ii) a comparative pass-rate analysis
against common post-processing methods under four bias regimes. All simulation scripts are publicly available in the Data and Code Availability statement.

%\subsection{Entropy convergence}
%Mudei aqui, tirei o itemize
%We simulated modular aggregation for four source models to %verify that the
%Shannon entropy of the output approaches $\log_2 %m$~\cite{CoverThomas2006}
%as $N$ increases (Figure~\ref{fig:entropy_plot}):

%\begin{itemize}
%\item \textbf{IID Strong Bias:} $m = 10$, $p_0 = 0.8$, %remaining
%probability uniform.
%\item \textbf{IID Extreme Bias:} $m = 21$, $p_0 = 0.9$, %remaining
%probability uniform.
%\item \textbf{Non-Stationary Random Noise:} $m = 3$; %probabilities
%drawn freshly at random at each step.
%\item \textbf{Non-Stationary Cyclic Bias:} $m = 5$; a dominant %outcome
%cycles periodically among the five values.
%\end{itemize}

%In all four cases the entropy converges rapidly to the %theoretical maximum
%$\log_2 m$ (dashed lines), regardless of the type of bias. %This entropy-based
%check is qualitative; the finite-sample comparison is reported %in the next
%subsection.

\subsection{Entropy convergence}

\begin{figure}[htbp]
\centering
\includegraphics[width=0.7\linewidth]{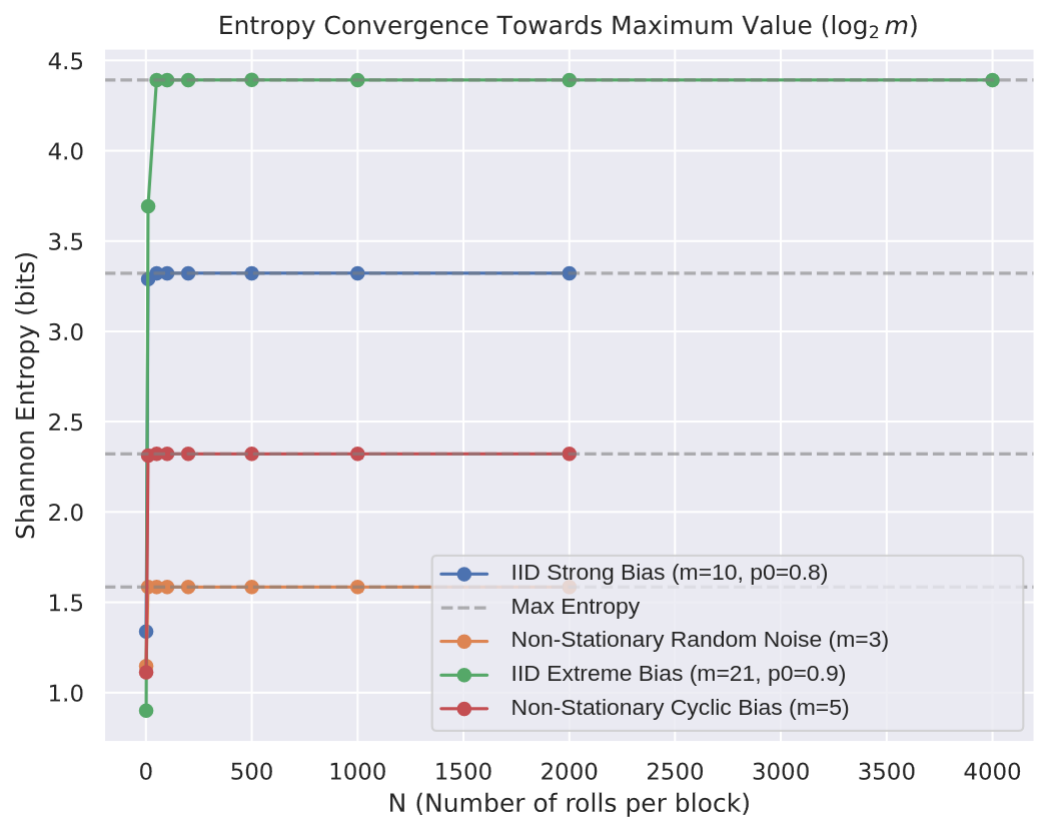}
\caption{Shannon entropy of the modular output as a function of the
aggregation size $N$ for four source models. Dashed lines indicate the
theoretical maximum $\log_2 m$ for each modulus.}
\label{fig:entropy_plot}
\end{figure}

We simulated modular aggregation for four source models to verify that the
Shannon entropy of the output approaches $\log_2 m$~\cite{CoverThomas2006}
as $N$ increases (Figure~\ref{fig:entropy_plot}): \textbf{(i) IID Strong
Bias:} $m = 10$, $p_0 = 0.8$, remaining probability uniform;
\textbf{(ii) IID Extreme Bias:} $m = 21$, $p_0 = 0.9$, remaining probability
uniform; \textbf{(iii) Non-Stationary Random Noise:} $m = 3$, probabilities
drawn freshly at random at each step; and \textbf{(iv) Non-Stationary Cyclic
Bias:} $m = 5$, a dominant outcome cycles periodically among the five values.

In all four cases the entropy converges rapidly to the theoretical maximum
$\log_2 m$ (dashed lines), regardless of the type of bias. This entropy-based
check is qualitative; the finite-sample comparison is reported in the next
subsection.

\subsection{Comparative analysis}

We benchmarked three methods---modular aggregation, the Peres
extractor~\cite{Peres1992}, and SHA-256~\cite{NIST2015}---under four source
scenarios with $m = 4$, evaluated against selected NIST SP 800-22
tests~\cite{Rukhin2010} (Frequency, Runs, Longest Run, Serial, DFT, and
Approximate Entropy tests). A run was counted as passing only if all p-values
exceeded $0.01$ \emph{and} the proportion of ones lay within
$[0.45,\,0.55]$. Each scenario was run $100$ independent times; the pass
counts below reflect those $100$ seeds.

The modular aggregation output is a symbol in $\mathbb{Z}_4$. For the binary
statistical tests only, each output symbol was encoded using the fixed map
$0\mapsto 00$, $1\mapsto 01$, $2\mapsto 10$, and $3\mapsto 11$. In all
experiments, successive modular outputs were produced from \emph{disjoint,
non-overlapping} blocks of raw samples, ensuring that the outputs themselves
are independent. In the non-stationary scenarios, independence between
successive draws was preserved, while the marginal distributions were allowed
to vary deterministically or randomly with time.

%\paragraph{Source scenarios}
%\begin{enumerate}
%\item \textbf{Static Asymmetric Bias (control):} Two outcomes each have
%probability 0.45; the remaining probability is split between the
%other two.
%\item \textbf{Random Additive Noise:} A base distribution is perturbed
%at each step by a random noise vector (noise level 0.3).
%\item \textbf{Moderate Cyclic Bias:} One dominant outcome ($p = 0.5$)
%whose index cycles through $\{0,1,2,3\}$.
%\item \textbf{Extreme Cyclic Bias (stress test):} Same structure, but
%dominant probability raised to 0.99.
%\end{enumerate}  %%change here

\paragraph{Source scenarios}
\textbf{(1) Static Asymmetric Bias (control):} Two outcomes each have
probability 0.45; the remaining probability is split between the
other two.
\textbf{(2) Random Additive Noise:} A base distribution is perturbed
at each step by a random noise vector (noise level 0.3).
\textbf{(3) Moderate Cyclic Bias:} One dominant outcome ($p = 0.5$)
whose index cycles through $\{0,1,2,3\}$.
\textbf{(4) Extreme Cyclic Bias (stress test):} Same structure, but
dominant probability raised to 0.99.

Note that for the cyclic bias scenarios, the probability distribution is a
cyclic shift of a base distribution. Consequently, the Fourier moduli
$|G_i(\omega^j)|$ are invariant under time shifts, meaning the spectral gap
parameter $\rho$ remains strictly constant across all steps.

The aggregation size $N$ was set so that the total-variation error bound
\eqref{eq:error_bound} fell below $2^{-64}$ for each scenario. The resulting
values were $(\rho,N)=(0.565685,79)$ for the static asymmetric source,
$(0.580000,83)$ for random additive noise, $(0.333333,41)$ for moderate
cyclic bias, and $(0.986667,3336)$ for extreme cyclic bias. The Peres extractor was implemented in a recursive 4-ary comparison form and
applied directly to the raw 4-ary symbols. SHA-256 was applied
to the packed bitstream (four 2-bit symbols per byte, 512-bit blocks).

\begin{table}[htbp]
\centering
\caption{Pass rates (passed runs / total runs) across four source scenarios.
Pass criterion: all selected p-values $\ge 0.01$ and bit balance within
$[0.45,\,0.55]$.}
\label{tab:comparison-final}
\begin{tabular}{l c c c c}
\toprule
& \multicolumn{4}{c}{\textbf{Scenario}} \\
\cmidrule(lr){2-5}
\textbf{Method} & \textbf{Static} & \textbf{Random} & \textbf{Moderate} & \textbf{Extreme} \\
& \textbf{Bias} & \textbf{Noise} & \textbf{Cyclic} & \textbf{Cyclic} \\
\midrule
Modular Aggregation & 97/100 & 95/100 & 90/100 & 92/100 \\
Peres Extractor     & 92/100 & 99/100 & 0/100 & 0/100 \\
SHA-256 Post-processing   & 95/100 & 93/100 & 89/100 & 0/100 \\
\bottomrule
\end{tabular}
\end{table}

Table~\ref{tab:comparison-final} shows distinct behaviours across the tested
procedures. The Peres extractor, designed for i.i.d.\ sources~\cite{Peres1992},
did not pass the selected statistical criteria under the cyclic-bias
scenarios considered here. SHA-256 handled moderate cyclic bias in most runs,
but did not pass the extreme stress-test configuration. SHA-256 was not
designed as a formal randomness extractor; its security guarantees assume
high-entropy inputs, a condition violated under extreme cyclic bias. The
observed failure is therefore consistent with known limitations of
cryptographic hash functions when used outside their design assumptions.
Modular aggregation achieved high pass rates across all tested scenarios,
with occasional finite-sample failures. This behaviour is consistent with the
exponential decay of each non-zero Fourier mode established in
Section~\ref{sec:theory}, which does not depend on the temporal structure of
the bias under the assumptions considered here.

\subsection{The security--retention trade-off}

\begin{figure}[htbp]
\centering
\includegraphics[width=0.7\linewidth]{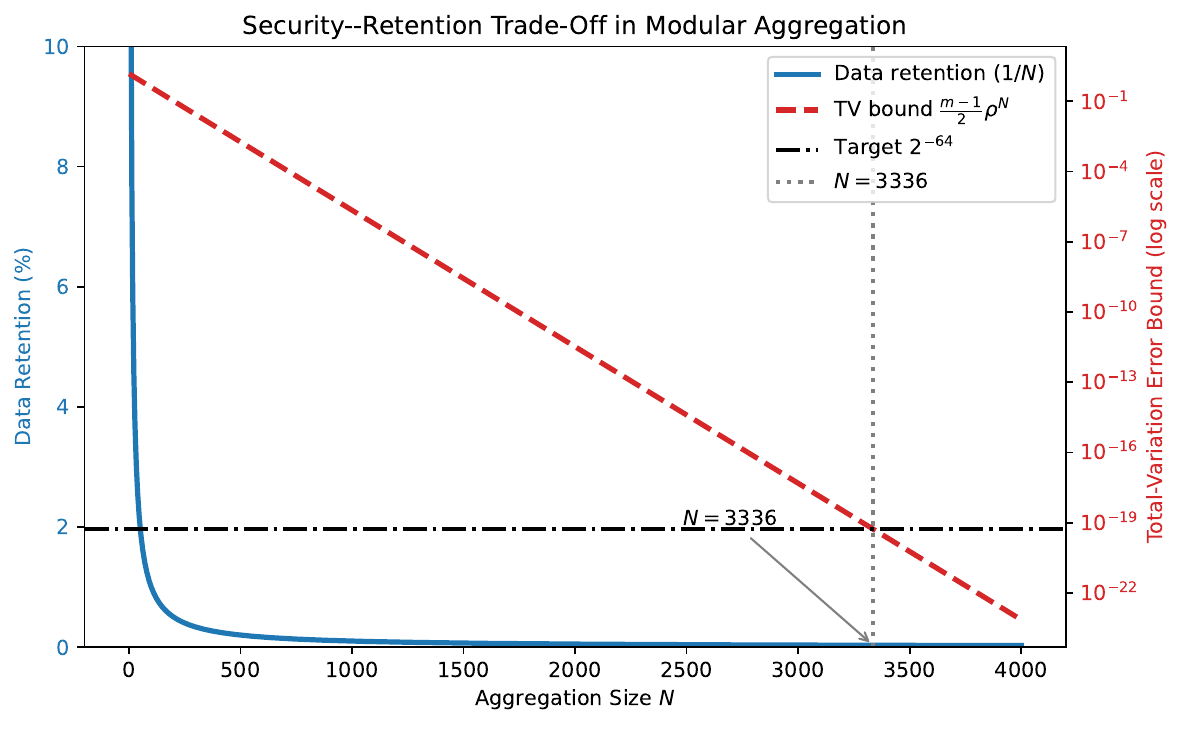}
\caption{Security--retention trade-off in modular aggregation. Data
retention ($\propto 1/N$, blue) and the total-variation error bound
($\tfrac{m-1}{2}\rho^N$, dashed red) are shown as functions of the
aggregation size $N$. The vertical dotted line marks $N = 3336$, the
operational point for the extreme cyclic-bias scenario.}
\label{fig:tradeoff_plot}
\end{figure}

Robustness comes at a cost. Each aggregation block of $N$ raw samples
produces a single output symbol, so the data retention rate scales as $1/N$.
Figure~\ref{fig:tradeoff_plot} shows this trade-off together with the
total-variation error bound $\tfrac{m-1}{2}\rho^N$. For the extreme cyclic
bias scenario ($\rho = 0.986667$), the threshold $N = 3336$ required to make
the bound fall below $2^{-64}$ is highlighted.

\section{Discussion and Conclusion}
\label{sec:conclusion}

We have shown that modular aggregation converges exponentially to the uniform distribution for independent discrete sources on $\mathbb{Z}_m$ under a uniform spectral-gap condition on the non-trivial Fourier modes. The rate of convergence is controlled by the largest non-trivial Fourier modulus $\rho$, yielding the explicit error bound in Eq.~\eqref{eq:error_bound}. Numerical simulations illustrate the theoretical predictions and indicate that, within the tested scenarios, modular aggregation remains robust against strong deterministic cyclic drift, unlike the tested implementations of the Peres extractor and SHA-256.

The method is mathematically transparent, unseeded, and its single free parameter $N$ provides a direct, tunable link between data retention rate ($\propto 1/N$) and statistical quality. Because it provides explicit convergence guarantees, it is well suited for generating the seed for a subsequent, faster extractor. Additionally, the Fourier criterion provides a useful diagnostic tool: if a non-trivial Fourier mode remains close to unit modulus, convergence becomes slow or may fail, indicating insufficient variation in some modular direction. Persistent failures may therefore suggest structural degradation of the source, although finite-sample effects, temporal correlations, encoding choices, and implementation details must also be taken into account.

Future work could extend the analysis to sources with intrinsic temporal correlations (e.g., Markov chains), derive non-asymptotic bounds, and explore experimental implementations with physical random-number generators such as photon-counting devices~\cite{marsili2013, HerreroCollantes2017}.

\subsection*{Data and Code Availability}
\label{sec:code}

The simulation scripts supporting this study are available at\\
\footnotesize{\url{https://colab.research.google.com/drive/1xqyWENOxHbBGbR5lR52UXCkJVCmMHoEF}}

\bibliographystyle{elsarticle-num}
\bibliography{citartigo}

\end{document}